\documentstyle[aps,epsfig,amsmath]{revtex}

\begin{document}

\title{Decay of correlations in the dissipative two-state system}
\author{
  G. Lang $^{1}$, 
  E. Paladino $^{2,1}$,  and
       U. Weiss $^{1}$           }
 \address{ 
   ${1}$ Institut f\"{u}r Theoretische Physik, Universit\"{a}t
           Stuttgart, D-70550 Stuttgart, Germany\\
    ${2}$   Istituto di Fisica, Universit\`a di Catania {\rm\&} INFM,
           Viale A. Doria 6, 95129 Catania, Italy.      }
   \maketitle

\begin{abstract}

We study the equilibrium correlation function of the polaron-dressed 
tunnelling operator in the dissipative two-state system and compare the
asymptoptic dynamics with that of the position correlations. 
For an Ohmic spectral density with the damping strength 
$K=\frac{1}{2}$, the cor\-relation functions are obtained in analytic form 
for all times at any $T$ and any bias.  
For $K<1$, the asymptotic dynamics  is found by using a diagrammatic approach
within a Coulomb gas representation. At $T=0$, the tunnelling or
coherence correlations drop as $t^{-2K}$, whereas 
the position correlations show universal decay $\propto t^{-2}$.
The former decay law is a signature of unscreened attractive charge-charge
interactions, while the latter is due to  unscreened dipole-dipole 
interactions.

\noindent 05.30.-d, 05.40.+j, 73.40.Gk\\
\end{abstract}

The simplest model that allows to study the interplay of tunnelling and 
dissipation is the spin-boson model~\cite{leggett,book}. Despite its simplicity, it 
exhibits generic features of many complex systems in physics and chemistry 
and has found widespread applications.
It has been adopted to describe diverse systems like the tunnelling of atoms 
in atomic-force devices \cite{micr}, or the 
dynamics of the magnetic flux in a rf-SQUID~\cite{squid}, 
just to mention a few.

The Ohmic spin-boson model exhibits a dynamical
phase transition between coherent and incoherent tunnelling. 
Both for the  expectation value $\langle\sigma_z(t)\rangle$ (the population) 
and the $\sigma_z$
autocorrelation function, the transition from
oscillatory to overdamped behaviour was found to occur at a damping strength
$K=\frac{1}{2}$ at $T=0$ and zero bias~\cite{egg-grab-w,lesage97,voelker}. 
Concerning the dynamics at long times, these functions behave quite
differently. The factorized system-reservoir initial state for 
 $\langle\sigma_z(t)\rangle$ leads to exponential
decay~\cite{egg-grab-w,lesage97}, whereas the correlated initial state of the
symmetrized equilibrium correlation function implies an algebraic decay
$\propto 1/t^2$~\cite{sass-w902,costi-ki}.

The dissipative two-state system (TSS) is described by the 
spin-boson Hamiltonian 
\begin{equation}
\label{H2}
H = H_0 + \sum_{\alpha} \bigg[\,\frac{p{}_\alpha^2}{2m_\alpha} + \frac{1}{2} 
        m_\alpha \omega_\alpha^2 \left(x_\alpha 
        - \frac{c_\alpha}{m_\alpha \omega{}_\alpha^2} \frac{a}{2} 
          \sigma_z \right)^2 \bigg]\;,
\end{equation}
where $H_0 = - \hbar \left( \Delta \sigma_x + \epsilon \sigma_z \right)/2$
describes the isolated TSS with level splitting $\hbar\Delta$ 
and a bias energy $\hbar\epsilon$. 
The  eigenstates of $\sigma_z$ are the two
localized states $|R\rangle$ and $|L\rangle$ at positions $\pm a/2$.
The effects of an Ohmic bath are captured by the spectral density 
$J(\omega) = ( 2 \pi \hbar  K / a^2)\,\omega\,e^{- \omega / \omega_c}$, where
$K$ is a dimensionless damping strength and $\omega_c$ is a cutoff
for the bath modes. Here we are interested in the scaling limit 
$\Delta_r/\omega_c\to 0$, where 
$\Delta_r = \Delta ( \Delta/\omega_c ) ^ {K/(1-K)}$ is kept finite.
In this limit, the population $\langle\sigma_z(t)\rangle$ and 
the $\sigma_z$ autocorrelation function are universal functions of
the renormalized frequency $\Delta_r$, i.e., there is no explicit dependence
on $\omega_c$.

Recently, there has been considerable interest for observables of the 
tunnelling operator $\sigma_x$ \cite{GWW,strong}. 
The correlations of $\sigma_x$ describe the response of the system to a 
change in the barrier width or in the barrier height. 
In atomic force devices, this can be contrived by modulating
the distance between surface and tip, and in the rf-SQUID
by varying the critical current \cite{squid}.
It was shown that, because of an explicit $\omega_c$ dependence,
both the expectation value and the autocorrelation function of the bare
$\sigma_x$ vanish  in the scaling limit \cite{guinea}.

In this Letter, we study the correlation function of the polaron-dressed
tunnelling operator $ \tilde{\sigma}_x =  U \sigma_x U^{-1}$ which has a 
nontrivial scaling limit.
The polaron transformation
$U = \exp \left\{- i \sigma_z\sum_{\alpha} s_\alpha p_\alpha /{2 \hbar} 
\right \}$ induces adiabatic displacements 
$\{s_\alpha \} \equiv \{ a c_\alpha/m_\alpha \omega{}_\alpha^2 \}$
of the bath modes during the tunnelling process 
(cf. Ref.~\cite{leggett,silb-har}).
The dressed tunnelling operator,
$\tilde\sigma_x = |R\rangle\langle L| \exp 
\{ -i \sum_{\alpha} s_\alpha p_\alpha /\hbar \}\, +\;{\rm h}.
\;{\rm c}.\;$, transfers the particle from one 
localized state to the other and simultaneously shifts the bath modes
by the displacements $\{\pm s_\alpha\}$. 
It is just in this way that coherent oscillations are induced and therefore
$\tilde\sigma_x$ is called the coherence operator.

The quantities of our interest are  the equilibrium correlation functions 
of the position, 
$C_z(t)=\langle\sigma_z(t)\sigma_z(0)\rangle_\beta
-\langle\sigma_z\rangle_\beta^2$, and of the coherence operator,
$C_x(t)=\langle\tilde\sigma_x(t)\tilde\sigma_x(0)\rangle_\beta$.
Our emphasis is put on their asymptotic behaviour for general $K<1$.
Using a real-time path integral approach, an exact formal series expression 
for the symmetrized position correlation function $S_z(t)=\mbox{Re}\,C_z(t)$
has been derived in Ref.~\cite{sass-w901},
\begin{align}
      S_z(t)&=1+\sum_{m=1}^{\infty} 
                 \left(-\bar{\Delta}^2\right )^{m}
          \int_{-\infty}^{t}\!{\mathcal D}^{(0)}_{0,2m}\{t_j\}
            \sum_{\{\xi_j=\pm 1\}} G_{0,m} D^{(+)}_{0,m}\;\,
            +Q_s(t)-\langle\sigma_z\rangle_\beta^2 \;,\label{ps}
\end{align}
\begin{align}
 Q_s(t)&=-\sum_{n=1}^{\infty}\sum_{m=1}^{\infty}
             \left(-\bar{\Delta}^2\right )^{n+m} \tan^2(\pi K)
             \int_{-\infty}^{t}\! {\mathcal D}^{(0)}_{2n,2m}\{t_j\}
 \sum_{\{\xi_j=\pm 1\}} \xi_1\,\xi_{n+1}\, G_{n,m} D^{(+)}_{n,m}\;.\label{qs}\\
  \langle\sigma_z\rangle_\beta&=-\lim_{t\to \infty}\sum_{m=1}^{\infty} 
        \left(-\bar{\Delta}^2\right )^{m} \tan(\pi K)
         \int_{-\infty}^{t}\! {\mathcal D}^{(0)}_{0,2m}\{t_j\}
        \sum_{\{\xi_j=\pm 1\}} \xi_1\, G_{0,m} D^{(-)}_{0,m}\;.
\end{align}
Similarly, the respective response function
$\chi_z(t) = -(2/\hbar)\,\Theta(t)\,{\rm Im}\,C_z(t)$ is found as 
\begin{align}
  \chi_z(t)&=-\frac{2}{\hbar}\sum_{n=1}^{\infty}\sum_{m=0}^{\infty} 
             \left(-\bar{\Delta}^2\right )^{n+m}
               \tan(\pi K)\int_{-\infty}^{t}\! {\mathcal D}^{(0)}_{2n-1,2m+1}
            \{t_j\} \sum_{\{\xi_j=\pm 1\}}
               \xi_1\,\xi_{n}\,G_{n,m} D^{(+)}_{n,m}\;.\label{chiz}
\end{align}
Here we have put $\bar\Delta^2 = \Delta^2\cos(\pi K)/2$.
The integration symbol contains the time-ordered integrations over
$k$ flips in the negative and $l$ flips in the positive time branch,
\begin{align}\label{intsymb}
  \int_{t_0}^t{\mathcal D}^{(r)}_{k,l}\{t_j\}&=
       \int_0^t\! dt_{k+l+r}\!\int_0^{t_{k+l+r}}\!
dt_{k+l+r-1}\,
           \cdots\!
       \int_0^{t_{k+2+r}}\!dt_{k+1+r}\int_{t_0}^0 dt_{k}\,
            \cdots
       \int_{t_0}^{t_2}\! dt_{1}\;.
\end{align}
In these expressions, the summations over the intermediate diagonal (sojourn)
states have already been performed. Further, the $\xi_j$ summation
takes into account the two possibilities for each off-diagonal (blip)
state the system takes. The effect of the bias is in  the factors 
$ D^{(+)}_{n,m}=\cos(\epsilon\sum_{j=1}^{n+m} \xi_j \tau_j)$  and
$ D^{(-)}_{n,m}=\sin(\epsilon\sum_{j=1}^{n+m} \xi_j \tau_j)$,
 where $\tau_j=t_{2j}-t_{2j-1}$ is the length of blip $j$.
The bath-induced correlations  are expressed in terms 
of the interaction $S_{p,q}= S(t_p-t_q)$, where 
$S(t)= 2K \ln[\,(\hbar\beta\omega_c/\pi)\sinh(\pi t/\hbar\beta)\,]$.
In the equivalent Coulomb gas picture, each factor $\exp[-S(t_p-t_q)]$  
represents the interaction of a neutral pair of charges, which we refer to as
a dipole. The full intra- and interblip correlations are in the factor
\begin{align} \label{G}
  G_{n,m}& = \exp \Big[ - \sum_{\substack{j=1}}^{n+m} S_{2j, 2j-1} -
                  \sum_{\substack{j=2}}^{n+m} \sum_{\substack{k=1}}^{j-1}
                  \xi_j \xi_k \Lambda_{j,k} \Big]\;,
\\       \label{Lambda}
  \Lambda_{j,k} &= S_{2j,2k-1} + S_{2j-1,2k} - S_{2j,2k} - S_{2j-1,2k-1}\;. 
\end{align}

For $C_x(t)$, the procedure is more complicated, since the polaron-dressed 
operator $\tilde\sigma_x$ acts on the full system-plus-reservoir space. 
Therefore, we have reconsidered the elimination of the bath modes. We have 
found that the resulting influence functional can be cast into the standard
Feynman-Vernon form at the expense of introducing modified system paths.
The findings can be put in simple terms:
the system's jumps at times zero and $t$ enforced by the operation of 
$\tilde\sigma_x$ do not give rise to bath correlations in the influence 
functional.

The contributions to $C_x(t)$ can be divided into two groups.
Either the system jumps from a sojourn to a blip
state at time zero (group A)  or vice versa (group B). This extra jump
is accounted for by the upper index $r=1$ in the integration element 
(\ref{intsymb}).
For both groups, the 
system is finally in a blip state. The symmetrized correlation
function  $S_x(t)=\mbox{Re}\,C_x(t)$ and the response function
$\chi_x(t)=-(2/\hbar)\,\Theta(t)\,\mbox{Im}\,C_x(t)$ are found as
\begin{align}
S_x^{\rm{A}}(t)  &= \frac{1}{2}\sum_{m=1}^{\infty} 
                 \left(-\bar{\Delta}^2\right )^{m-1}
          \int_{-\infty}^{t}\!\!{\mathcal D}^{(1)}_{0,2m-2}\{t_j\}
            \sum_{\{\xi_j=\pm 1\}^{\rm{A}}}
                G_{0,m}^{\rm{A}} D^{(+)}_{0,m}  \label{sxa} \; , \\
S_x^{\rm{B}}(t) &= -\sum_{n=1}^{\infty}\sum_{m=2}^{\infty}
             \left(-\bar{\Delta}^2\right )^{n+m-1} \sin^2(\pi K)
             \int_{-\infty}^{t}\!\! {\mathcal D}^{(1)}_{2n-1,2m-1}\{t_j\}
   \!\!\!\!\! \!\!\!\!\sum_{\substack{\{\xi_j=\pm 1\}\\ 
                       \xi_n=\xi_{n+1}=-\xi_{n+m}}} \!\!\!\!\!\!\!\!\! 
                \xi_1\,\xi_{n+m}\, G_{n,m}^{\rm{B}} D^{(+)}_{n,m} \;, 
\label{sxb}\\ 
\chi_x^{\rm{A}}(t)   &= \frac{1}{\hbar}\sum_{n=1}^{\infty}\sum_{m=1}^{\infty} 
             \left(-\bar{\Delta}^2\right )^{n+m-1}
               \tan(\pi K)\int_{-\infty}^{t}\!\! {\mathcal D}^{(1)}_{2n,2m-2}
            \{t_j\}  \sum_{\{\xi_j=\pm 1\}^{\rm{A}}} 
               \xi_1\,\xi_{n+m}\,G_{n,m}^{\rm{A}}D^{(+)}_{n,m}\;,\label{axa} \\
\chi_x^{\rm{B}}(t) &= \frac{1}{\hbar}\sum_{n=1}^{\infty}\sum_{m=1}^{\infty} 
              \left(-\bar{\Delta}^2\right )^{n+m-1}
                  \tan(\pi K) \int_{-\infty}^{t} \!\! 
                   {\mathcal D}^{(1)}_{2n-1,2m-1}\{t_j\}   \notag  \\ 
            &\qquad \times \sum_{\{\xi_j=\pm 1\}^{\rm{B}}} 
               \xi_1 \, G_{n,m}^{\rm{B}} D^{(+)}_{n,m} 
            \;\{\sin^2(\pi K) \, \xi_{n+1} + \cos^2(\pi K) \, \xi_{n+m} \}\;. 
\label{axb}
\end{align}
 The superscripts 
$\{ \ldots \}^{\rm{A}}$
and  $\{ \ldots \}^{\rm{B}}$ indicate that the blip labels are subject to
the constraints
\begin{align}\label{cnstr}
\xi_{n+m} &=\xi_{n+1} \qquad\mbox{(group A)}\;, &
\xi_{n+m} &=-\xi_{n} \qquad \mbox{(group B)}\;.
\end{align} 
The interaction factors $G^{\rm A}_{n,m}$ and  $G^{\rm B}_{n,m}$
differ from the form (\ref{G}) for
$G_{n,m}$ by the removal of the charges at times  $t'=0$ and 
$t'=t$. We have
\begin{align}
\label{GA}
  G^{\rm{A}}_{n,m}& = \exp \Big[ - \sum_{j=1\;,\;j\neq n+1}^{n+m} 
                S_{2j, 2j-1} -
             \sum_{\substack{j=2}}^{n+m} \sum_{\substack{k=1}}^{j-1}
               \xi_j \xi_k \Lambda^{\rm{A}}_{j,k} \Big] \;,\\
    G^{\rm{B}}_{n,m}& =\exp\Big[-\sum_{j=1\;,\;j\neq n}^{n+m} 
                  S_{2j, 2j-1} -
            \sum_{\substack{j=2}}^{n+m} \sum_{\substack{k=1}}^{j-1} 
             \xi_j \xi_k \Lambda^{\rm{B}}_{j,k} \Big] \;,
\end{align}
where $\Lambda^{\rm{A}}_{j,k}$ and $\Lambda^{\rm{B}}_{j,k}$ describe the 
interblip correlations for the modified sequences of charges. 
If $j,\,k\neq n+1$ and $\neq n+m$, $\Lambda^{\rm{A}}_{j,k}$ is
 again given by  (\ref{Lambda}). In all other cases, the interactions of the
 missing charges in Eq.~(\ref{Lambda}) have to be dropped. For instance, for
 $j=n+1$, we have $\Lambda^{\rm{A}}_{n+1,k}=S_{2n+2,2k-1}-S_{2n+2,2k}$.
There are analogous modifications in $\Lambda^{\rm{B}}_{j,k}$ for $j,\,k=n$ 
and $n+m$. For instance, we have 
$\Lambda^{\rm{B}}_{n,k}=S_{2n-1,2k}-S_{2n-1,2k-1}$.

\begin{figure}[t]
\setlength{\unitlength}{0.55cm}
\centerline{
\begin{picture}(22,5)
\put(-1.08,0.5){$-\infty$}
\put(4,0.5){$0$}
\put(8.5,0.5){$t$}
\put(13.42,0.5){$-\infty$}
\put(18.5,0.5){$0$}
\put(23.0,0.5){$t$}
\put(-0.5,4.27){\epsfig{file=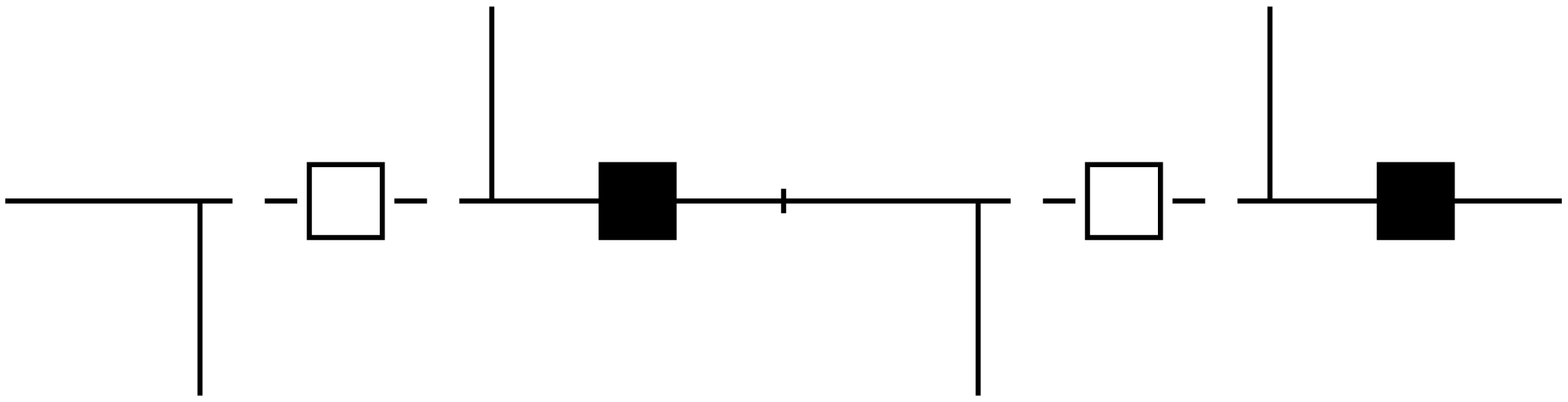,width=2.36cm,angle=-90}} 
\put(14.0,3.8){\epsfig{file=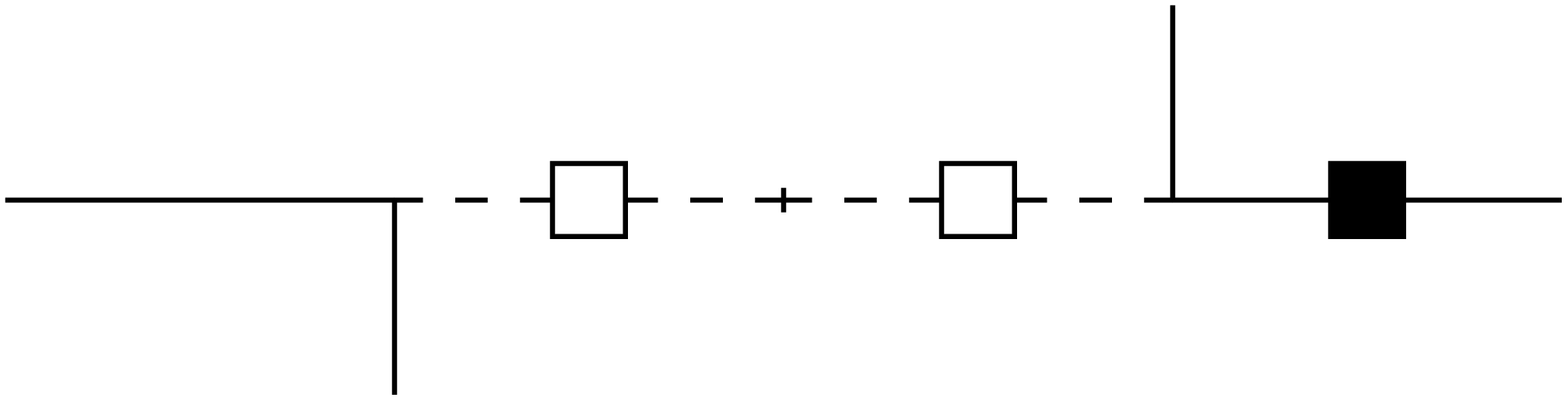,width=1.87cm,angle=-90}}
\end{picture}}
\caption{\small The diagrams for $\chi_z(t)$ (right) and the contribution 
$Q_s(t)$ to $S_z(t)$ (left). The full and dashed lines 
represent sojourns and blips, respectively. An empty box  represents the
insertion of a CS gas within a blip interval, the full box stands for a 
CB gas inside a sojourn interval. The upward and downward spikes
symbolize charges. \label{figurez}} 
\end{figure}

For the value $K=\frac{1}{2}$, the above series  
can be summed in analytic form using the concept of collapsed dipoles
\cite{sass-w901}. 
Putting $K=\frac{1}{2}-\kappa$ with $\kappa \ll 1$, 
the phase factor  
$\cos(\pi K)\approx \pi\kappa$ vanishes in the limit 
$\kappa\rightarrow 0$.
In order to have a finite contribution for $K=\frac{1}{2}$, each factor 
$\cos(\pi K)$
has to be compensated by  the $1/\kappa$ ``short-distance'' singularity 
arising from the breathing mode integral of a dipole with interaction
$ e^{-S(\tau)} \approx (\omega_c \tau)^{-(1-2\kappa)}$. 
The combined expression is termed a collapsed dipole, yielding the finite
contribution
\begin{equation}
\label{gamma}
I(K={\textstyle \frac{1}{2}})=
\lim_{K\rightarrow  1/2}\,\Delta^2 \cos(\pi K)
     \int_0 d\tau \, e^{-S(\tau)} \, 
=\frac{\pi}{2}\frac{\Delta^2}{\omega_c} \equiv \gamma  \quad .
\end{equation}
A collapsed dipole does not interact with other charges and it 
is insensitive to a symmetric bias factor.
In contrast, an odd bias factor in Eq.~(\ref{gamma}) prevents a 
dipole from collapsing. However, to give a nonzero contribution, the 
extended dipole must be free of a $\cos(\pi K)$ factor.
Within an extended blip [\,sojourn\,] state of length $\tau$ [\,$s$\,], 
the system 
may make any number of 
visits of duration zero to a sojourn [\,blip\,] state. This 
is described
by the insertion of a grand-canonical gas of non-interacting collapsed 
sojourns (CS) [\,or blips (CB)\,], yielding a factor $e^{-\gamma \tau/2}_{}$
[\,or  $e^{-\gamma s}_{}$\,]. Note that there is a multiplicity factor 2
for collapsed blips.
As a general rule, an extended sojourn interval, say $s_k$, is free of 
insertions only if the
subsequent blip is weighted with a factor $\xi_{k+1}$. In the expression 
(\ref{qs}), {\it e.g.}, the first sojourns in each time branch are
without insertions.
Employing these concepts, the following results for the various
correlation functions are obtained. The $\sigma_z$ correlation functions are
given by \cite{book} 
\begin{align}\label{sz12}
 S_z(t)&=e^{-\gamma t}-F_+^2(t)-F_-^2(t)\;,\qquad\qquad
\chi_z(t)=(4/\hbar)\, F_+(t)\,e^{-\gamma t/2} \;, \\ 
F_{\pm}(t)& =\frac{\Delta^2}{2\gamma}\int_0^{\infty}\! d\tau\,e^{-S(\tau)}
      f_{\pm}(\epsilon\tau)
        \big( e^{-\gamma|t-\tau|/2}\mp e^{-\gamma(t+\tau)/2}\big)\;, 
\end{align}
with $ f_{+}(x)=\cos(x)$,   $ f_{-}(x)=\sin(x)$.
A diagrammatic representation of $S_z(t)$ and $\chi_z(t)$ is displayed in 
Fig.~\ref{figurez}.
\vspace{0.25cm}
\noindent In the asymptotic regime $T=0$ and  $t\gg 1/\gamma$, we find from 
Eq.~(\ref{sz12})
\begin{equation}
  S_z(t)=-\left(\frac{4}{\pi\gamma}\right)^2\left(\frac{\gamma^2}{
\gamma^2+4\epsilon^2}\right)^2\, \frac{1}{t^2} \;,\qquad
\chi_z(t) =\frac{16}{\pi\hbar}\frac{\gamma^2}{
\gamma^2+4\epsilon^2}\, 
\frac{\cos(\epsilon t)\,e^{-\gamma t/2}}{\gamma t} \;.
\end{equation}
The exponential decay of the response function is due to
the grand-canonical sums of collapsed dipoles in
each interval (except for the first sojourn). 
In contrast, the function $S_z^{}(t)$ 
decays algebraically because the first sojourn in the positive time 
branch is free of collapsed blips. Therefore, this interval 
gets very large and is effectively limited by the overall 
length $t$. The $1/t^2$ law 
reflects the bare interaction between the two dipoles displayed in 
Fig.~\ref{figurez} (left).

\begin{figure}[t]
\setlength{\unitlength}{0.7cm}
\centerline{
\begin{picture}(9.5,2.5)
\put(0,0.5){$-\infty$}
\put(5,0.5){$0$}
\put(9.5,0.5){$t$}
\put(0.5,1.7){\epsfig{file=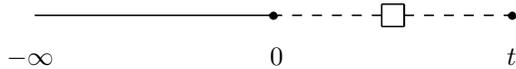,width=0.353cm,angle=-90}}
\end{picture}}
\caption{\small The diagram describing $S_x(t)$.
The bullets mark transitions which are free of bath correlations 
because of the modified influence functional.  
\label{figuresx}}
\end{figure}

Now switch to the $\tilde\sigma_x$ correlation functions.  Since there is
one $\cos(\pi K)$ factor more than dipoles in Eq.~(\ref{sxb}), 
$S_x^{\rm{B}}(t)$
vanishes as $K\to \frac{1}{2}$. Therefore, only group A contributes to
the symmetrized correlation function, yielding the damped 
oscillatory behaviour
\begin{align}
  S_x(t)&=\cos(\epsilon t)\,e^{-\gamma t/2}\;.\label{sx12}
\end{align}
The corresponding diagram is sketched in Fig.~\ref{figuresx}. 
\begin{figure}[t]
\setlength{\unitlength}{0.55cm}
\centerline{
\begin{picture}(22,5)
\put(-1.08,0.5){$-\infty$}
\put(4,0.5){$0$}
\put(8.5,0.5){$t$}
\put(13.42,0.5){$-\infty$}
\put(18.5,0.5){$0$}
\put(23,0.5){$t$}
\put(-0.5,4.27){\epsfig{file=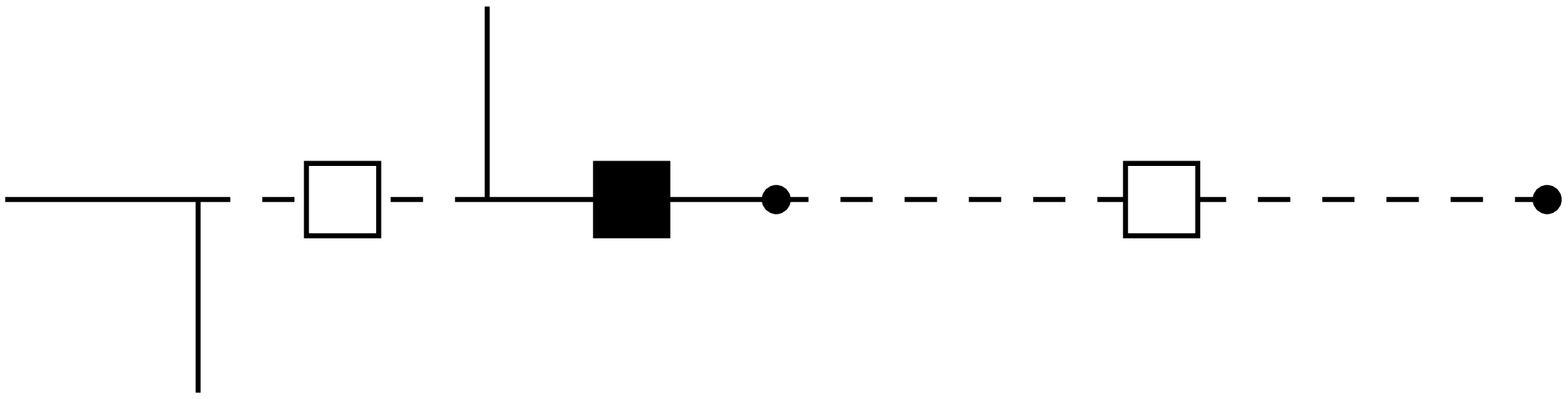,width=2.02cm,angle=-90}}
\put(14,3.8){\epsfig{file=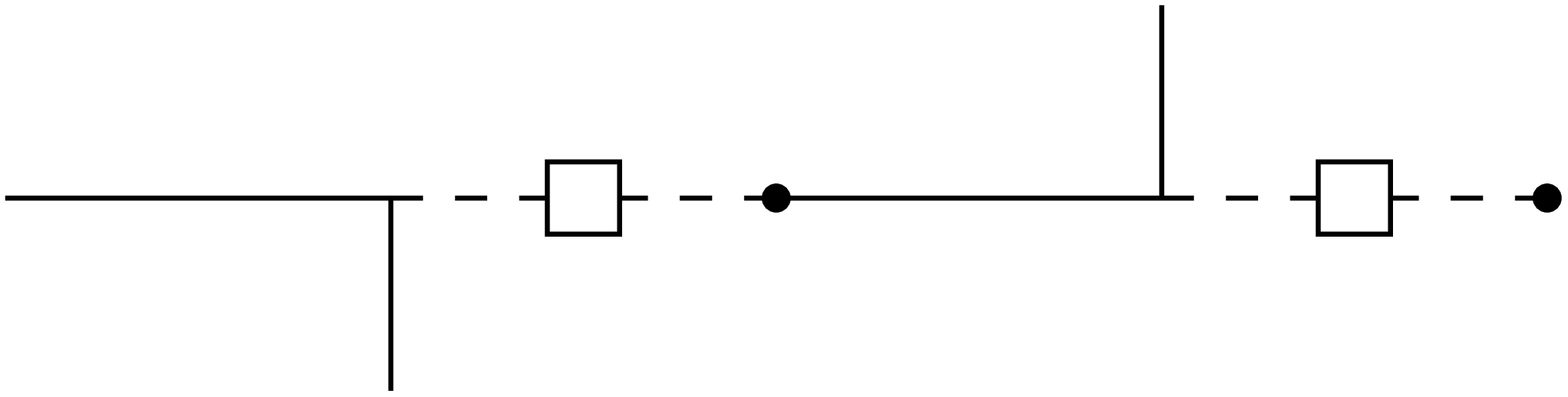,width=1.87cm,angle=-90}}
\end{picture}}
\caption{\small The contribution of group A (left) and  
group B (right) to $\chi_x(t)$.
Each diagram has only one extended dipole.\label{figurechix}}
\end{figure}
\vspace{0.25cm}
\noindent For the response function, the contributions of both groups 
are shown in Fig.~\ref{figurechix}. The expressions are combined to 
\begin{align}\label{chix12}
\chi_x(t)&=
(2/\hbar) \big[\cos(\epsilon t) F_+(t) + \sin(\epsilon t) F_-(t) \big]\;.
\end{align}
At $T=0$ and times $t\gg 1/\gamma$, we find from Eq.~(\ref{chix12})
\begin{equation} \label{chi12long}
\chi_x(t)=\frac{8}{\pi\hbar}\frac{\gamma^2}{
\gamma^2+4\epsilon^2}\, \frac{1}{\gamma t} \;.
\end{equation}
The algebraic decay law arises from the contribution of group B.
According to the above rule, the  sojourn interval  in 
the positive time branch is free of a CB gas factor and therefore 
is effectively very large for $t\gg 1/\gamma$.
The $1/t$ law in Eq.~(\ref{chi12long}) is simply the signature of the bare
intra-dipole interaction, $e^{-S(t)}\propto 1/t$ for $K=\frac{1}{2}$.

For general $K$, it is not possible to sum the series for the 
$\sigma_z$ and $\tilde\sigma_x$ correlation functions in analytic form.
Nevertheless, it is possible to extract the long-time behaviour of the 
correlations. 
The essential modifications concern the CB and CS gas factors inserted in a 
given interval. For $K\neq \frac{1}{2}$, the dipoles are no longer 
collapsed and thus the respective grand-canonical sum can not
be performed in analytic form. However, because of the alternating sum,
the charges which are partitioned off by the weight factors $\xi_{k+1}$
form clusters of effective length $1/\Delta_r$.

Consider first the $\sigma_z$ correlation functions at asymptotic times
$t\gg 1/\Delta_r$. For $\chi_z(t)$,
there is a single neutral cluster surrounding the origin of the
time axis. Therefore $\chi_z(t)$ decays exponentially. 
In $S_z(t)$, we have a neutral cluster in each time branch.
Since in both branches the initial sojourn is free of insertions
(see above rule), the two clusters are near the origin and near $t$,
respectively.
The correlations between the two neutral clusters  are dominated by the bare
dipole-dipole interaction which is $-2K/t^2$ at $T=0$. The 
interaction has a universal 
power-law form which is independent of the coupling strength. 
The power $2$ is a characteristic feature of Ohmic dissipation. 
The dipole moment emerges to coincide with the static
susceptibility \cite{sass-w902}. In the end, we have
\begin{align}\label{szlong}
  S_z(t)&=-2K \,\big[\,\hbar\chi_z^{(0)}/2\,\big]^2\,\frac{1}{t^2} \;.
\end{align}
In the frequency domain, Eq.~(\ref{szlong}) corresponds to the generalized
Shiba relation \cite{sass-w902} for the spectral function,\linebreak
$S_z(\omega\to 0)=2\pi K\,\big[\,\hbar\chi_z^{(0)}/2\,\big]^2\,|\omega|$.

Next, consider the $\tilde\sigma_x$ correlation function for $t\gg 1/\Delta_r$.
In group A, there is only a single neutral cluster near to the origin of the
time axis. Hence, both
$S_x^{\rm{A}}(t)$ and $\chi_x^{\rm{A}}(t)$ decay exponentially at 
asymptotic times. In group B, we have a charged cluster in each time branch,
satisfying overall neutrality. Again, as a result of the above rule,
the clusters are situated near the origin and near $t$ and therefore
are roughly separated by an interval of length $t$.
The clusters interact with unscreened charge-charge interaction 
$e^{-S(t)} \propto t^{-2K}$. This interaction directly determines 
the long-time behaviour of $S_x^{\rm B}(t)$ and $\chi_x^{\rm B}(t)$.
The contributions of group B predominate over the exponential contributions
of group A for $t\gg 1/\Delta_r$. Thus we find the asymptotic behaviour
\begin{align}\label{asbev}
S_x (t) &\propto e^{-S(t)}_{} \propto t^{-2K} \;,&
\chi_x(t) &\propto e^{-S(t)}_{} \propto t^{-2K} \;.
\end{align} 
We note that the prefactor in the algebraic decay law for $S_x(t)$
vanishes accidentally as $K\to \frac{1}{2}$. The results (\ref{asbev}) have 
been indicated numerically in Ref.~\cite{strong}.
The slow decay of $\chi_x(t)$ for $K < \frac{1}{2}$ implies that the static
susceptibility diverges algebraically, $\propto T^{2K-1}$. Interestingly,
this regime coincides with the coherence regime for the 
population $\langle\sigma_z(t)\rangle$ at zero bias \cite{egg-grab-w,strong}.

In conclusion, we have given exact formal expressions for the tunnelling or
coherence correlations in the dissipative two-state system for general
damping strength $K$ and we have presented results in analytic form for the
particular case $K=\frac{1}{2}$. The differences in the long-time behaviours 
of the position and coherence correlations
have been illustrated in terms of a simple charge picture. 

\acknowledgments
E.P. thanks R. Fazio and G. Giaquinta for useful discussions and 
acknowledges financial support by the INFM under the PRA-QTMD programme. 
Partial support was provided by the Deutsche Forschungsgemeinschaft (DFG).

\end{document}